\documentclass[twocolumn]{article}

\usepackage{subcaption}
\usepackage{graphicx}
\usepackage{color}
\usepackage{soul} 
\usepackage{multirow}
\usepackage{appendix}
\usepackage{subcaption} 
\usepackage{ragged2e}
\usepackage[hyphens]{url}
\usepackage{booktabs}
\usepackage{amsmath}
\usepackage{amssymb}
\usepackage{inputenc}
\usepackage{array}
\usepackage{epsfig}
\usepackage{amsfonts}
\usepackage{geometry}
\usepackage[multiple]{footmisc}

\DeclareUnicodeCharacter{2212}{-}

\newcolumntype{M}[1]{>{\centering\arraybackslash}m{#1}}

\providecommand{\keywords}[1]
{
  \small	
  \textbf{\textit{Keywords---}} #1
}

\long\def\@maketitlenotetext#1#2{\noindent
            \hbox to 1.8em{\hss$^{#1}$}#2}

\newcommand*\samethanks[1][\value{footnote}]{\footnotemark[#1]}

\title{\textbf{Modelling Cooperation and Competition in Urban Retail Ecosystems with Complex Network Metrics}}

\author{
    \textbf{Jordan Cambe}\thanks{These two authors contributed equally and are listed alphabetically.} \textsuperscript{}\thanks{Work was performed while a visitor at the University of Cambridge. Complete affiliation:
    Univ Lyon, ENS de Lyon, UCB Lyon 1, CNRS, Laboratoire de Physique, F-69342 Lyon, France;
    Institut Rh\^{o}nalpin des Systemes Complexes, IXXI, F-69342 Lyon, France}\\
    \texttt{Laboratoire de Physique,}\\
    \texttt{ENS de Lyon}\\
    \texttt{jordan.cambe@gmail.com}

\and
    \textbf{Krittika D'Silva}\samethanks[1]\\
    \texttt{Computer Laboratory,}\\
    \texttt{University of Cambridge}\\
    \texttt{krittika.dsilva@cl.cam.ac.uk}

\and
    \textbf{Anastasios Noulas} \\ 
    \texttt{New York University}\\
    \texttt{tnoulas@gmail.com}

\and   
    \textbf{Cecilia Mascolo} \\
    \texttt{Computer Laboratory,}\\ 
    \texttt{University of Cambridge}\\
    \texttt{cm542@cam.ac.uk}

\and
    \textbf{Adam Waksman} \\
    \texttt{Foursquare Labs}\\
    \texttt{awaksman@foursquare.com}
}

\date{}

\begin{document}

\maketitle

\begin{abstract}
Understanding the impact that a new business has on the local market ecosystem is a challenging task as it is multifaceted in nature. Past work in this space has examined the collaborative or competitive role of homogeneous venue types (i.e. the impact of a new bookstore on existing bookstores). However, these prior works have been limited in their scope and explanatory power. To better measure retail performance in a modern city, a model should consider a number of factors that interact synchronously. This paper is the first which considers the multifaceted types of interactions that occur in urban cities when examining the impact of new businesses. We first present a modeling framework which examines the role of new businesses in their respective local areas. Using a longitudinal dataset from location technology platform Foursquare, we model new venue impact across 26 major cities worldwide. Representing cities as connected networks of venues, we quantify their structure and characterise their dynamics over time. We note a strong community structure emerging in these retail networks, an observation that highlights the interplay of cooperative and competitive forces that emerge in local ecosystems of retail establishments. We next devise a data-driven metric that captures the first-order correlation on the impact of a new venue on retailers within its vicinity accounting for both homogeneous and heterogeneous interactions between venue types. Lastly, we build a supervised machine learning model to predict the impact of a given new venue on its local retail ecosystem. Using two classes of features that account for the presence of different venue types as well as how these venue form cooperative networks, the model achieves area under the curve above 80\% scores for certain
venue categories. Our approach highlights the power of complex network measures in building machine learning prediction models. These models have numerous applications within the retail sector. More broadly, however, the methodology and results can support policymakers, business owners, and urban planners in the development of models to characterize and predict changes in urban settings.
\end{abstract}


\section{Introduction}
Location is known to be highly influential in the success of a new business opening in a city.  
Where a business is positioned across the urban plane not only determines its reach by clienteles of relevant demographics, but more critically, it determines its exposure to a local ecosystem of businesses who strive to increase their own share in a local market. 
The types of businesses and brands that are present in an urban neighborhood in particular has been shown \cite{jensen2} to play a vital role in determining whether a new retail facility will grow and blossom, or instead whether it will become a sterile investment and eventually close. Competition is nonetheless only one determinant in retail success. How a local business establishes a \textit{cooperative network} with other places in its vicinity has been shown to also play a decisive role in its sales growth \cite{daggitt}. Local businesses can complement each other by exchanging customer flows with regards to activities that succeed each other (e.g. going to a bar after dining at a restaurant), or through the formation of urban enclaves of similar local businesses that give rise to characteristic identities that then become recognisable by urban dwellers. A classic example of the latter is the presence of many Chinese restaurants in a Chinatown~\cite{zhou2010chinatown}.

It is therefore natural to hypothesize that the rise of a business lies on the complex interplay between cooperation and competition that manifests in a local area. Measuring these cooperative and competitive forces in a city remains, however, a major challenge. Today's cities change rapidly driven by urban migration and phenomena such as gentrification as well as large urban development projects, which can lead to shops opening and closing at increasing rates. In 2011, the \textit{fail rate} of restaurants in certain cities, such as New York, was as high as 80\% \footnote{\url{https://www.businessinsider.com/new-york-restaurants-fail-rate-2011-8}} with some businesses closing in only a matter of months. 
A similar picture has been reported for high street retailers in the United Kingdom with part of the crisis being attributed to the increasing dominance of online retailers
\footnote{\url{https://www.theguardian.com/cities/ng-interactive/2019/jan/30/high-street-crisis-town-centres-lose-8-of-shops-in-five-years}}. 

Data generated in location technology platforms by mobile users who navigate the city provides a unique opportunity to respond to the aforementioned challenges. In addition to providing quick updates, in almost real time, on the places that open (or close) in cities - thus accurately reflecting the set of local businesses in a given area - they offer a view on urban mobility flows between areas and places at fine spatial and temporal scales. The ability to describe these two dimensions of urban activity - places and mobility - paves the way for measuring the impact, either positive or negative, that retail facilities have on each another. In this work, we harness this opportunity, building on a longitudinal dataset by Foursquare that describes mobility interactions between places in 26 cities around the world. 
Our contributions are summarized in more detail in the following: 
\begin{itemize}
\item{\textbf{Detecting patterns of cooperation in urban activity networks:} We model businesses in a city as a connected network of nodes belonging to different activity types. We examine the properties of these networks spatially and temporally. With respect to a null network model, we observe higher clustering coefficient, higher modularity and lower closeness centrality scores which are indicators of strong tendencies for local businesses to cluster and form collaborative communities that exchange customer flows. 
In numerical terms, the modularity of urban activity networks is $\approx 0.6$  relative to the corresponding null models with $\approx 0.15$.}

\item{\textbf{Measuring the impact of new businesses using spatio-temporal metrics:}
We next model the impact of a new business opening in a given area. 
Previous work \cite{daggitt} conducted a preliminary analysis of homogeneous impact of new businesses. However, this work was limited as it did not isolate numerous contributing factors such as whether multiple new venues have opened when attributing the impact to a given business. We develop a methodology more robust to bias through the use of spatio-temporal filters Moreover our approach generalizes to heterogeneous interactions between venue types present in an urban system by considering impact measurements between different venue categories (e.g. measuring the impact of a restaurant to a bar). Notably, we observe that the opening of a Fast Food Restaurant near other Fast Food Restaurants results in the most significant competition, with a median decline in customer flows of 21\% over 6 months. We also show how this competitive ranking can be created for heterogeneous categories.}

\item{\textbf{Predicting optimal retail environment using urban dynamics and network topology measures:} Lastly, we built a supervised learning model to predict the impact of multiple new venues on an existing venue. We incorporate additional network metrics into our model to consider the urban environment of a given venue and its capacity to operate cooperatively with other places in vicinity. 
We observe significant heterogeneity between categories where some have more predictable trends while others have greater variations. In light of this heterogeneity across venue types we tailor supervised learning models by training them in a manner that reflects the idiosyncrasies of its category. Despite the inherent difficulty of the prediction task, due to the multi-factorial nature of dynamic and complex interactions in retail ecosystems, our results suggest that incorporating complex signals in predictive machine learning frameworks can offer meaningful insight in real world application scenarios. For certain business categories, AUC scores above 0.7 are consistently attained, whereas complex network metrics consistently boost the performance of classifiers offering a clear advantage over baseline methods.}
\end{itemize}

Our results are especially important in a digital age with shifting customer preferences as physical business are forced to adapt to remain competitive. Our methodology can enable a better understanding of interactions within local retail ecosystems.
Modern data and methods, such as those employed in the present work, not only can allow for monitoring these phenomena at scale, but also offer novel opportunities for retail facility owners to assess the risk of opening a new venture through location-based analytics. Similar methods can be applied beyond the scope of the retail sector we study here, namely for urban planning and innovation e.g. through assessing the impact of opening transport hubs, leisure and social centers or health and sanitation facilities in city neighborhoods.  

\section{Related Work}
Understanding retail ecosystems and determining the optimal location for a business to open have been long been questions in operations research and spatial economics~\cite{ghosh1983formulating,eiselt1989competitive}. Compared to modern approaches, these methods were characterized by static datasets informing on population distribution across geographies, tracked through census surveys and the extraction of retail catchment areas through spatial optimization methods~\cite{applebaum1966methods}.
Gravity models on population location and mobility later became a common approach for site placement of new brands~\cite{gibson1972retail}.


The availability of spatio-temporally granular urban datasets and the popularization of spatial analysis methods in the past decade led to a new generation of approaches to quantify retail success in cities. In this line, network-based approaches have been proposed to understand the retail survival of local businesses through quality assessment on the interactions of urban activities locally~\cite{jensen2}. In addition to networks of places, street network analysis emerged as an alternative medium to understand customer flows in cities, with various network centrality being proposed as a proxy to understand urban economic activities~\cite{crucitti2006centrality,porta2012street}. While this previous work examines the impact of new businesses, it is limited in scope. It considers only homogeneous categories, does not compare network properties across different cities, and does not build a prediction model. This is where the primary novelty of the present paper lies. 

More recently, machine learning and optimization methods have been introduced to solve location optimization problems in the urban domain, focusing not only on retail store optimization~\cite{karamshuk2013geo} but also real estate ranking~\cite{fu2014sparse} amongst other applications. Location technology platforms such as Foursquare opened the window of opportunity for customer mobility patterns to be studied at fine spatio-temporal scales~\cite{d'silva1, d'silva2} and moreover, semantic annotations on places presented direct knowledge on the types of urban activities that emerge geographically and led to works that allowed for the tracking and comparing urban growth patterns at global scale~\cite{daggitt}. Closer to the spirit of the present work from a modelling perspective, the authors in~\cite{hidalgo} study co-location patterns of urban activities in Boston and subsequently recommend areas where certain types of activities may be missing. 

\section{Dataset Description}
Within the last decade, Online Location-based Services have experienced a surge in popularity, attracting hundreds of millions of users worldwide. These systems have created troves of data which describe, at a fine spatio-temporal granularity, the ways in which users visit different businesses and areas of a city. We hypothesize these data can be used to build a predictive model of the impact of a new venue on the surrounding businesses. To this end, we utilize data from Foursquare, a location technology platform with a consumer application that allows users to check into different locations. As of August 2015, Foursquare had more than 50 million active users and more than 10 billion check-ins ~\cite{foursquarenumbers}.

The basis of our analysis is a longitudinal dataset from 26 cities that spans three years, from 2011 to 2013, and included over 80 million checkins. We aggregate data from the 10 most represented cities in North America\footnote{North American cities are: Austin, Boston, Dallas, San Francisco, New York City, Houston, Las Vegas, Los Angeles, Toronto, Washington.} and the 16 most represented cities in Europe\footnote{European cities are: Amsterdam, Antwerpen, Barcelona, Berlin, Brussels, Budapest, Copenhagen, Gent, Helsinki, Kiev, Madrid, Milano, Paris, Prague, Riga, London}. A summary of our data is described in Table~\ref{tab:data}. 

For each venue, we have the following information: geographic coordinates, specific and general category, creation date, total number of check-ins, and number of unique visitors. The specific and general categories fall within Foursquare’s API of hierarchical categories. A full list of the categories can be found by querying the Foursquare API \cite{venueapi}.
The dataset also contains a list of transitions within a given city. A transition is defined as a pair of check-ins by an anonymous user to two different venues within the span of three hours. It is identified by a start time, end time, source venue, and destination venue.
 We consider the set of venues $V$ in a city. A venue $v \in V$ is represented with a tuple $<loc, date, category>$ where $loc$ is the geographic coordinates of the venue, $date$ is its creation date, and $category$ is the specific category of the venue.
The creation date, $date$, for a given venue refers to the date it was added to the Foursquare platform. 
Prior work by Daggitt et al. \cite{daggitt} showed that across all cities when examining the number of venues added per month, the last 20\% of venues were new venues rather than existing venues added to the database for the first time. We apply this methodology to all cities and define new venues as those that fall within the last 20\% of venues added to Foursquare for that city.

\begin{table} 
\begin{center}
\begin{tabular}{c|>{\centering\arraybackslash}m{1.4cm}|>{\centering\arraybackslash}m{1.4cm}|>{\centering\arraybackslash}m{1.6cm}}
\hline
 Region & \# venues & \# new venues & \# transitions \\ \hline
 North America & 94,094 & 29,552 & 43,200,432 \\ \hline 
 Europe & 101,101 & 40,275 & 44,600,446 \\ \hline
 Total & 195,195 & 69,827 & 87,800,878 \\
 \hline
\end{tabular} 
\caption{Foursquare dataset description for 10 North American cities and 16 European cities.}
\label{tab:data}
\end{center}
\end{table}

\begin{figure*}
\centering
\begin{minipage}{.5\textwidth}
  \centering
  \includegraphics[width=8.9cm]{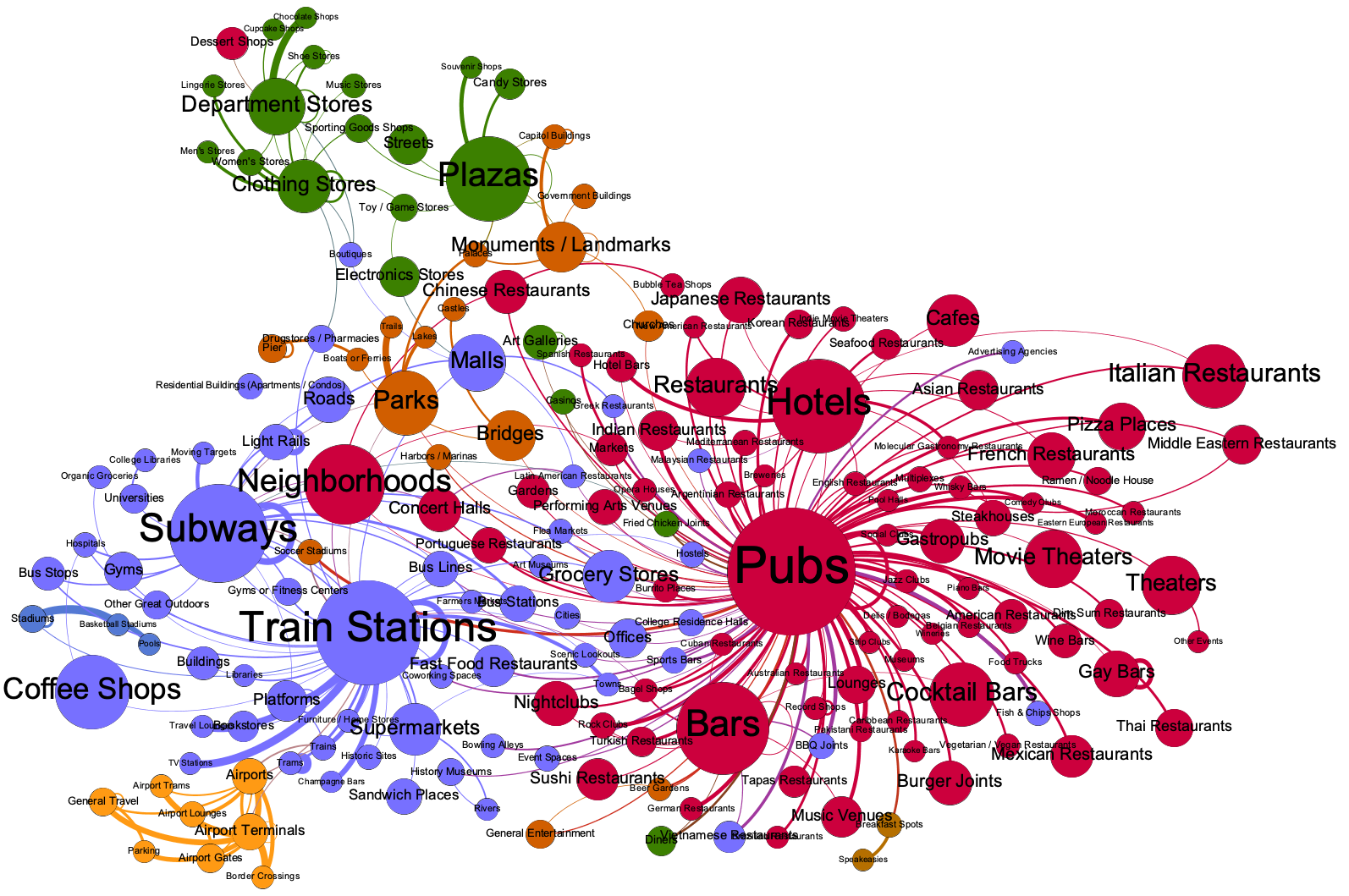}  
\end{minipage}%
\begin{minipage}{.5\textwidth}
  \centering
  \includegraphics[width=8.9cm]{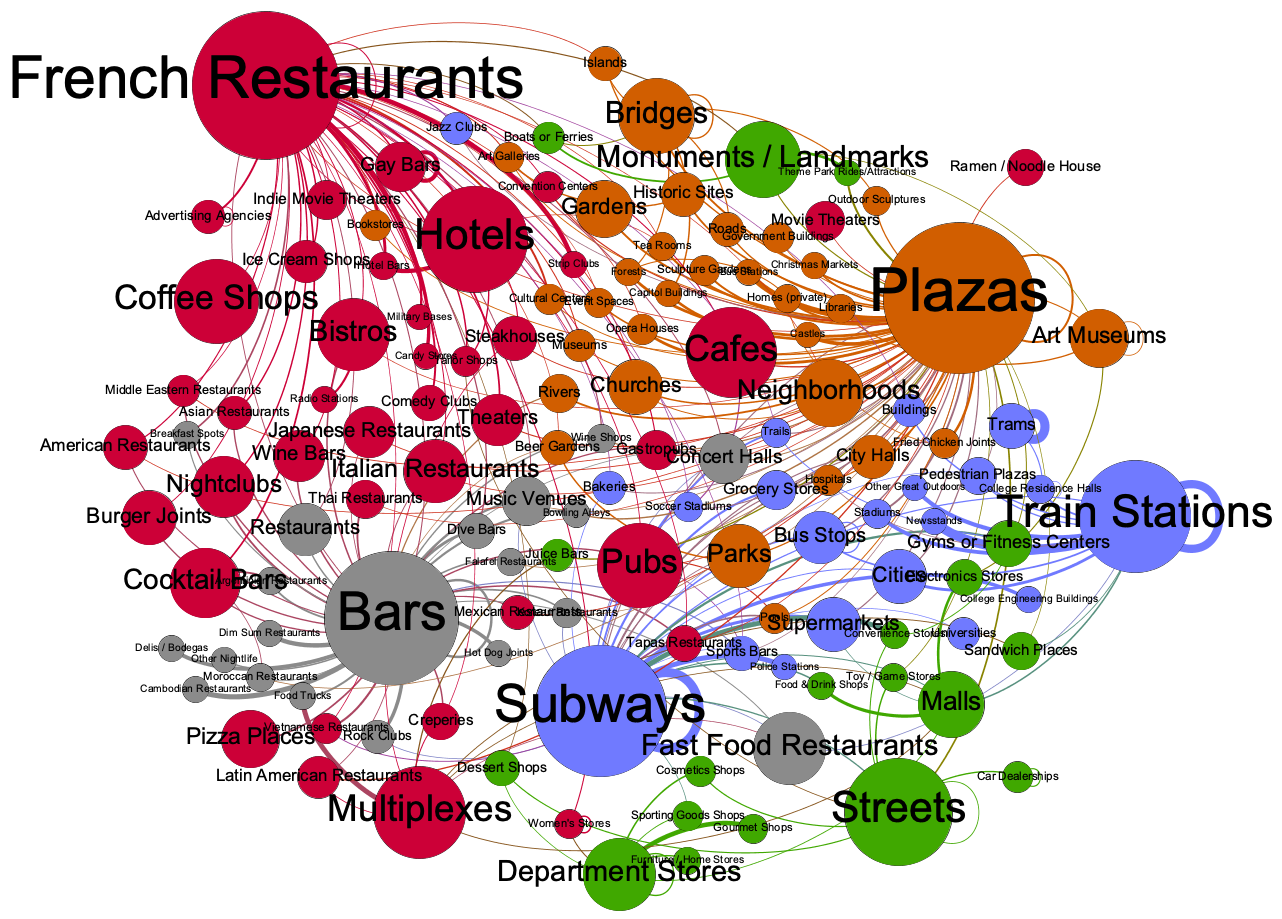}
\end{minipage}
    \caption{Network visualization of categories during the evening in London (left) and Paris (right). Different colors represent different Louvain communities ~\cite{blondel_2008}. We see clusters of travel and transport (blue), nightlife (red), and shopping activities (green).}
    \label{fig:networkViz}
\end{figure*}
\label{sec:data}

\section{Urban Activity Networks}
\label{sec:network}
We begin by examining transitions between Foursquare venues of different category types that we refer to as urban activities. While we are considering a mix of categories users check in in the city, our focus from an analysis and modelling point of view will be focusing on urban activities corresponding to retail establishments (e.g. restaurants). 

\subsection{Visualizing Mobility Interactions}
To visualize an urban activity network, we create a graph $G_i$ for each city $i$, where the set of nodes $N_{cat}$ is the set of business categories defined previously in Section \ref{sec:data}. In this network, business categories are linked by weighted directed edges $e_{s\rightarrow d}$. A directed link is created from the source category $c_s$ to the destination category $c_d$ if at least one transition happens during the time window we consider (e.g. weekend, weekday, or a period of hours during a day). Thus, the weight of each edge is proportional to the total number of transitions from the source category to the destination category for the particular time period of interest for each city. The weights are then normalized by the total number of check-ins that occurred at $c_d$. Therefore, the weight can be interpreted as the percentage of customers of $c_d$ who come from $c_s$. To eliminate insignificant links, we filter out edges that have less than 50 transitions total. We examine two time intervals of interest: morning AM (6am-12pm) and evening PM (6pm-12am). 

In Figure~\ref{fig:networkViz} we visualize the network in the evening for two cities, London and Paris. The colors represent different communities, obtained using the Louvain community detection algorithm~\cite{blondel_2008}. Further, the size of nodes is proportional to their degree. This visualization, as one example, describes similarities and variations in the structure of urban activities in different cities.  We observe an underlying common structure for the two cities, even though cultural distinctions can also be noted. We have observed a similar pattern across different cities which we don't visualize due to lack of space. In terms of similarity in network structure, we see a shopping cluster (green) centered around Department Stores;  a cluster for travel and transport (blue) centered around categories such as Train Stations and Subways; a leisure cluster (light brown) centered around Plazas and containing outdoor categories (e.g. Parks, Gardens, Soccer Stadiums). On the other hand, differences in network structure become also apparent. We note for instance how recreation activities in the evenings differs across the two cities. London has a considerably large nightlife cluster (red) centered around pubs from which a number of different nightlife categories unfold (e.g. Nightclubs, restaurants of different types, Theater). Paris is more segregated and contains two nightlife clusters: one cluster around French Restaurants (red) linked to Coffee Shops, Theaters, Nightclubs; and another cluster (gray) centered around Bars which contains Food Trucks, Fast Food Restaurants, and Music Venues. This dichotomy translates to the presence of two classes of customers each of which adheres to different types of activity sequences during nighttime. Another observation is regarding variations in network structure over time: the Coffee Shop category in London is separated from the nightlife cluster, which may indicate different kind of customer behaviors between daytime and evening. Interestingly, we also see associations emerging between types of businesses. Taking Paris as an example, French Restaurants interact a lot with Coffee Shops and Nightclubs and so do Bars with Food Trucks. In both cities, Coffee Shops are drawing crowds from Subways, Toy Stores with Electronics Stores and Sport Stores.
 
Overall, these results suggest strong structural characteristics in urban activity networks where different categories of places form interaction patterns of cooperation, where mobile users move from one to the other. Competition on the other hand manifests in a more implicit manner in the network in two ways: first, retail facilities that are grouped in the same node (e.g. Bars) have to share customers that have been previously performing a different activity (e.g. going to a Restaurant) and second, through activities that do not share an edge in the network and as a result they do not interact with one another in terms of mobility patterns. 

\subsection{Network Properties}

\begin{table*}
\centering
\captionsetup{justification=centering}
\begin{tabular}{>{\centering\arraybackslash}m{1.6cm}|>{\centering\arraybackslash}m{1.1cm}>{\centering\arraybackslash}m{1.1cm}>{\centering\arraybackslash}m{1.1cm}>{\centering\arraybackslash}m{1.1cm}|>{\centering\arraybackslash}m{1.1cm}>{\centering\arraybackslash}m{1.1cm}>{\centering\arraybackslash}m{1.1cm}>{\centering\arraybackslash}m{1.1cm}}
    \toprule 
 \multicolumn{5}{c}{\qquad\qquad\qquad\quad \textbf{London}}     & \multicolumn{4}{c}{\textbf{Paris}}      \\  \toprule 
& AM & Random AM & PM & Random PM & AM & Random AM & PM & Random PM \\ \hline
\# of nodes & 176 & 176 & 199  & 199 & 157 & 157 & 150 & 150  \\ \hline
\# of edges & 1727 & 1125 & 2539 & 1636 & 1708 & 1084 & 1818 & 1105  \\ \hline
$<C>$  & 0.655   &  0.361         & 0.657   & 0.410  & 0.671   &      0.456     &  0.701  &   0.437       \\ \hline
$<C_c>$ & 0.298   &  0.426         &   0.373 &   0.444        &  0.367  &   0.445        & 0.390    & 0.454          \\ \hline
$Q$ &  0.380  &   0.172        &  0.398  &    0.153       & 0.326   &  0.156         & 0.310   &   0.147        \\ 
\bottomrule
\end{tabular} 
\caption{Network metrics for London and Paris for during the morning AM (6am - 12pm) and evening PM (6pm-12am). These metrics are compared to a configuration model (Random).}
\label{tab:networkMetrics}
\end{table*}

We next quantify the structure of these networks in terms of different network properties considering also different time intervals. For our two cities of comparison, we list the network metrics in Table~\ref{tab:networkMetrics} and enlist those next.\\

\begin{itemize}
    \item \emph{The average clustering coefficient}, $<C>$, is the tendency of categories to form triangles, that is to gather locally into fully connected groups. It varies between 0 and 1 with higher values implying a higher number of triangles in the network (see \cite{newman_2010} for more details). 
    \item \emph{The average closeness centrality}, $<C_c>$, is the average length of the shortest path between the nodes and here accounts for the tendency of categories to be close to each in terms of shortest paths \cite{newman_2010}. It varies between 0 and 1 where a higher closeness centrality score for a node suggests higher proximity to other nodes in the network.
    \item \emph{The modularity}, $Q$, is a well established metric indicating how well defined communities are within the network \cite{blondel_2008}. Modularity values fall within the range $[-1, 1]$, with greater positive values indicating greater presence of community structure.
\end{itemize}  

We compare our network metrics to a random baseline, a configuration model\cite{molloy_1995}, which maintains the degree distributions of our networks, in Table \ref{tab:networkMetrics}. The comparison with the null model provides an indication of how significant empirical observations are with respect to the random case. First we note that for all three metrics the real networks are very different to the corresponding null models. 
In general, high clustering coefficient and modularity together with a lower closeness centrality scores point to the tendency of local businesses to form significantly tight clusters that are well isolated from one another. Furthermore, we also investigated variations of these networks properties for different period of the day. We observed in some cases that the closeness centrality was higher in the evening relative to morning hours, which is the case of London for example (see Table \ref{tab:networkMetrics}).

Looking closer at the network modularity scores presented in Table \ref{tab:networkMetrics} we note a partitioning of different categories into communities with scores around $0.3/0.4$ for both cities compared to much smaller values $\approx0.15$ for the null model.
Finally, the similarity in terms of network properties values between the two cities, as well as the prominent community structure in both suggest that the hypothesis that the organization of the retail business ecosystem is similar across cities is a plausible one. This is true to a certain degree, nonetheless variations are also noted due to apparent cultural differences.

In this section, we highlighted the dominance of community structure and local clustering in urban activity networks. This observation suggests that categories gain (or lose) attractiveness as a result of other activities around them. It further raises the question of how businesses affect each other and in particular how these dynamics alter when a new business opens in proximity of another one. In the next section we perform an impact analysis with regards to opening events of new retail facilities aiming to shed light to the aforementioned questions.

\section{Measuring Impact} 
\label{sec:firstdegree}
In this section we examine the impact of new businesses on other establishments within their vicinity. Previous work~\cite{daggitt} has considered the \textit{homogeneous} impact of a new business opening, that is the impact that venue categories have on categories of the same type (e.g. the impact of a new Coffee Shop on another Coffee Shop). In this section we generalize the impact metric to heterogeneous mixes of categories, and develop a methodology more robust to spatio-temporal bias effects. 

\subsection{Spatio-temporal Scope of Impact} 
To measure the impact of a new venue opening in an area we need to define its geographic scope. We define the spatial neighborhood of a venue as the set of venues that are located within the radius $r_s$. Formally, we define the spatial neighbourhood of a venue $v_n$ as:
\begin{equation}
SN(v_n, r_s) = \{v_e \in V : dist(v_n, v_e) < r_s \wedge v_n \neq v_e\}
\end{equation}
where $V$ is the set of venues in the city and $dist(v_n, v_e)$ is the Euclidean distance between venue $v_n$ and $v_e$.

Further we introduce a similar notion across the temporal dimension. In particular, given a temporal radius $r_t$, then two businesses opening within $r_t$ are considered to be \emph{temporal neighbors}. Formally, we define the temporal neighbors of a venue $v_n$ as: 
\begin{equation}
TN(v_n, r_t) = \{v_e \in V: t\_dist(v_n, v_e) < r_t \wedge v_n \neq v_e\}
\label{eq:temporalNeigh} 
\end{equation}
where $t\_dist$ is the difference in the number of months between the creation date of $v_n$ and $v_e$. 
Finally, we define $W_T$ as the temporal window of observation, i.e. the total period in months over which we measure the number of check-ins at a given business.

\subsection{Measuring Impact}
\subsubsection{Defining the impact formula}
\label{sec:impact}
We base our impact metric $I_{v_n}(v_e,t_{v_n})$ of a new venue $v_n$ on an existing venue $v_e$ on the metric introduced in Dagitt et al. \cite{daggitt}. This is defined as the normalized number of transitions for an existing venue in a specific time period prior to and after a new venue opens in its spatial neighborhood. We define the number of check-ins during the posterior time interval as follows: 
\begin{equation}
C_{v_n}^{post}(v_e, t_{v_n}, \Delta t) = \sum_{d=t_{v_n}}^{t_{v_n}+\Delta t-1} n_{v_e}\left(d,d+1 \right)
\end{equation}
This calculates the sum of check-ins at venue $v_e$, $n_{v_e}$, after the opening of $v_n$ at $t_{v_n}$ and over a period of $\Delta t = W_T/2$ months. We define the number of checkins during the prior time interval as follows: 
\begin{equation}
C_{v_n}^{prior}(v_e, t_{v_n}, \Delta t) = \sum_{d=t_{v_n}}^{t_{v_n}-\Delta t-1} n_{v_e}\left(d-1,d \right)
\end{equation}

Similarly, this calculates the number of check-ins over the period of $\Delta t$ months before the opening of $v_n$. Formally, the impact metric is defined as follows:
\begin{equation}
I_{v_n}(v_e, t_{v_n}, \Delta t) =  
\frac{C_{v_n}^{post}(v_e, t_{v_n}, \Delta t)  / N_{cat}(t_{v_n},t_{v_n}+\Delta t)}
{C_{v_n}^{prior}(v_e, t_{v_n}, \Delta t)  / N_{cat}(t_{v_n},t_{v_n}-\Delta t) }
\label{eq:impact}
\end{equation} 
$N_{cat}(t_{v_n},t_{v_n}+\Delta t)$ (and respectively $N_{cat}(t_{v_n},t_{v_n}-\Delta t)$) is the total number of check-ins to all businesses of $v_e$'s category during the period $[t_{v_n},t_{v_n}+\Delta t]$ (respectively $[t_{v_n},t_{v_n}-\Delta t]$). This normalizing factor takes into account both the background trend of the category and the potential season effect which can occur in the market. This factor is calculated at a per city level.

\begin{table}
\centering
\captionsetup{justification=centering}
\begin{tabular}{>{\centering\arraybackslash}m{1.8cm}|>{\centering\arraybackslash}m{1.5cm}|>{\centering\arraybackslash}m{1.5cm}|>{\centering\arraybackslash}m{1.5cm}}
\hline
Category & Median Impact & \% Businesses $I<1$  & \# Businesses \\ \hline
Fast Food Restaurants & 0.79 & 67.9 & 28 \\ \hline
Bakeries & 0.81 & 73.1 & 26 \\ \hline
Pizza Places & 0.81 & 69.3 & 39 \\ \hline
Coffee Shops & 0.84 & 66.4 & 202 \\ \hline
Sandwich Places & 0.84 & 62.0 & 63 \\ \hline 
\end{tabular}    
\caption{Ranking of the categories which have the strongest homogeneous negative impact.}
\label{tab:competitiveRanking}
\end{table}

\begin{figure}
\centering
\subfloat[Effect of the size of the spatial radius on the number of data points when $r_t$ is set to 3 months. An optimal value of $r_s$ is drawn for 100 m.]{%
  \includegraphics[width=0.45\textwidth]{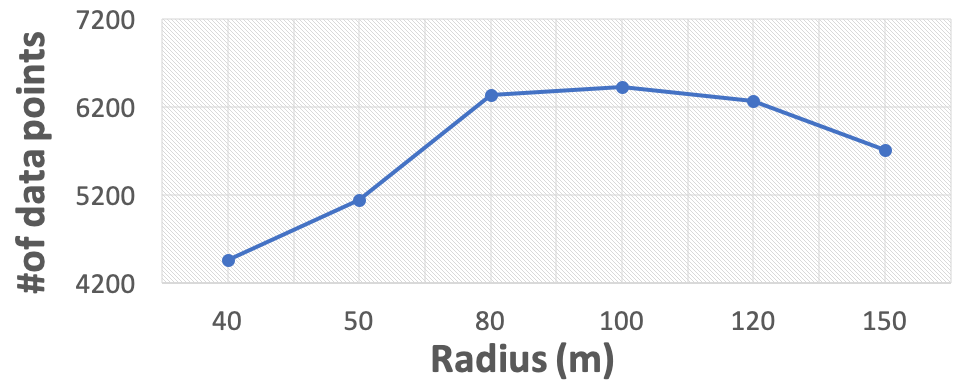}%
  }\par
\subfloat[Effect of the size of the temporal radius on the number of data points when $r_s$ is set to 100m. The number of data points continues to decrease for $r_t \geqslant 1$ months.]{%
  \includegraphics[width=0.45\textwidth]{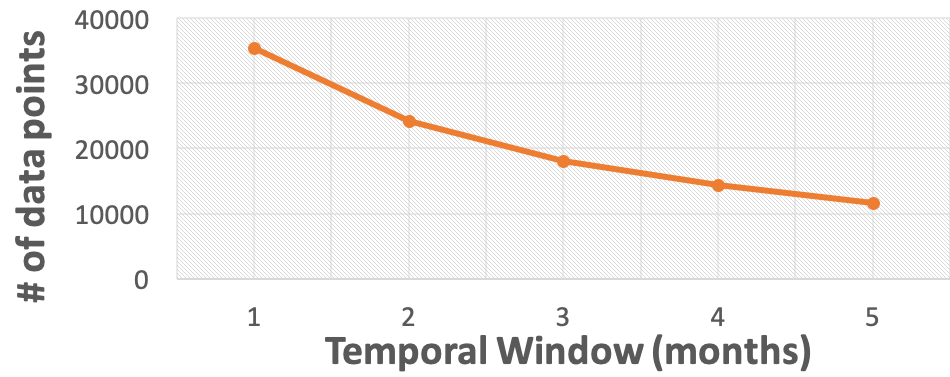}%
  }\par   
\caption{Tuning the spatial and temporal parameters of our model.}
\label{fig:tunparams}
\end{figure}

\subsubsection{Impact metric interpretation} 
Above, we define $N_{cat}(t_{v_n},t_{v_n}+\Delta t)$ as the total number of check-ins to all businesses of a specific category. We define the \emph{market share} of a given venue as the total number of check-ins to that venue as a percentage of $N_{cat}$. Using the impact metric defined in Section~\ref{sec:impact} we can calculate the percentage of market share gained or lost by venue $v_e$ after the opening of venue $v_n$. For example, if $v_n$ is a clothing store and $v_e$ is a sandwich shop $I_{v_n}(v_e, t_{v_n}, \Delta t) = 1.2$ means the market share of venue $v_e$ (the sandwich shop) increased by 20\% over the $\Delta t$ months following the opening of $v_n$ (the clothing store). 
Conversely, a number below 1 means the market share of the venue $v_e$ decreased after the opening of venue $v_n$, suggesting the new venue was a competitor. For the example above if $I_{v_n}(v_e, t_{v_n}, \Delta t) = 0.81$ this would correspond to the market share of venue $v_e$ (the sandwich shop) decreasing by 19\% over the $\Delta t$ months following the opening of $v_n$ (the clothing store).

\subsection{Tuning Spatial and Temporal Windows}
A major challenge in analyzing correlations in a complex system is implementing a methodology which can limit the effect of potential hidden variables which may bias the observed correlations.

For our model, our methodology must isolate the impact of \emph{one business} in constantly changing neighborhoods. For this reason, as a first analysis, we set $r_t$ to $\Delta t = W_T / 2$, that is half of the window of observation, and we only choose a pair $(v_n, v_e)$ if the new business $v_n$ has no temporal neighbors (defined above in Equation~\ref{eq:temporalNeigh}). In other words, if we set $\Delta t = W_T / 2 = 3$ months, we only measure the impact of $v_n$ on $v_e$ if $v_e$ was open at least 3 months before $v_n$ and if no other business opened in the spatial neighborhood of $v_e$ during $[t_{v_n}-\Delta t, t_{v_n}+\Delta t]$. We apply this form of filtering to isolate the confounding effect of multiple businesses opening next to each other. This naturally introduces a trade-off with regards to the number of data points considered for our measurement. 

\begin{figure} 
\includegraphics[width=\columnwidth]{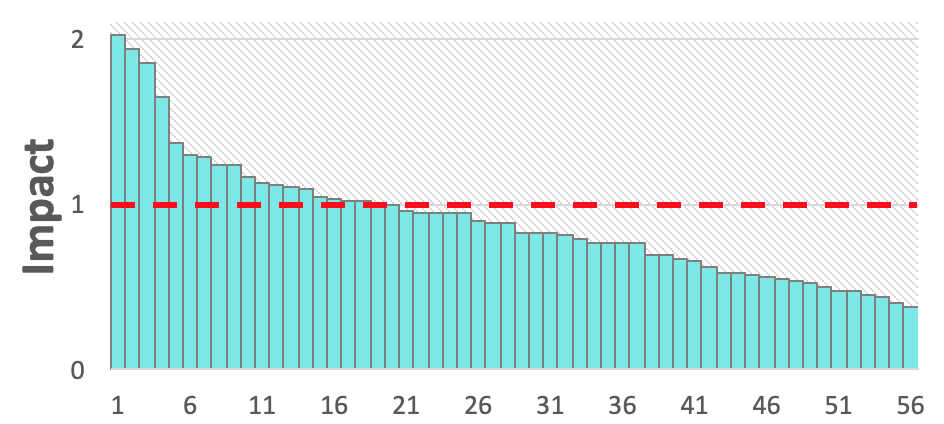}
\caption{Plot of the impact measure of Coffee Shops on Burger Joints. Each column is a Burger Joint for which only a Coffee Shop opened in its neighborhood during a period $\Delta t = 3$ months. We see that $37 / 56$ joints are over $I = 1$, that is 66\% of Burger Joints have been impacted negatively by the opening of a Coffee Shop.}
\label{fig:impact_individual}
\end{figure}
The number of data points resulting from our approach depends on two parameters. The spatial radius $r_s$ and the temporal radius $r_t$. As shown in Figure \ref{fig:tunparams}(a) where $r_t$ is fixed at 3 months, when the spatial radius is small the probability of a new business nearby opening is low. Hence the number of data points remains small as well. As the spatial radius increases so does the number of data points until a maximum is reached after which it decreases again. This likely due to the fact that after a certain spatial radius the probability of finding only one business opening within $r_s$ decreases as well. Based on these results we set parameter $r_s$ to $100$ meters for our subsequent analysis.
\begin{figure*} 
\begin{center}
\includegraphics[width=16cm]{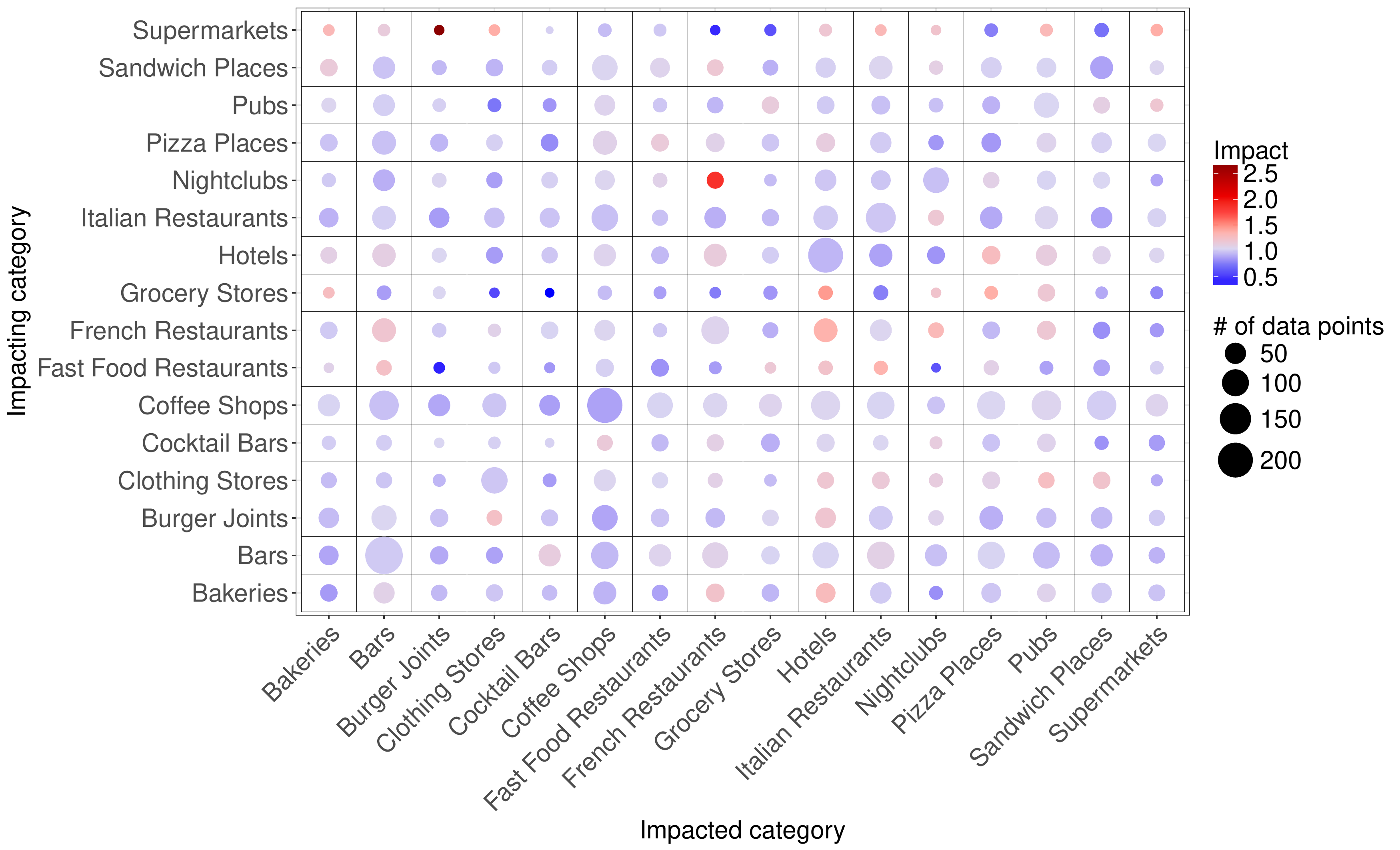}
\caption{Matrix of median impact of categories on each other. The size of the circles represent the number of pairs of businesses which were used (e.g. Coffee Shops - Coffee Shops is 200). The color varies according to the median impact value of the pair, dark blue corresponds to impact below 1 (representing competition), grey represents neutral median impact equal to 1, and dark red translates to high median impact (representing cooperation). Read: French Restaurants (9th row) have a median impact on Bars (2nd column) of 1.2 and less than 100 pairs were counted.}
\label{fig:impact_matrix}
\end{center}
\end{figure*} 
We apply the same reasoning for tuning the temporal radius $r_t$ setting $r_s$ to 100m. However, tuning $r_t$ is more challenging. Our hypothesis is that after a time period, $t_{v_n} + r_t$, the positive or negative impact of a new business will become stable within the scope of a neighborhood. After this period of time, $v_n$ will be considered as part of the baseline of the neighborhood, when examining the impact of future new venues. Our hypothesis is that as $r_t$ increases, the less likely we are to be considering other new shops that are interfering with the impact we measure. Figure \ref{fig:tunparams}(b) shows that the number of data points steadily decreases from $r_t = \Delta t > 1$ month onwards. This implies that the optimal value for $r_t$ in terms of data points is smaller than one month. However, as suggested earlier increasing $r_t$ limits the risk to consider previously new venues that opened just before the time window of interest in the impact measure. For this reason we choose a value of $r_t$ longer than one month. For the analysis described below, we set $r_t$ to three months.
 
\subsection{Measuring Impact on Retail Activity}
We perform the analysis using the aforementioned setup on the $16$ most popular retail categories and aggregated data from the 26 cities listed above in Section~\ref{sec:data}. We aggregated data from these cities based on observed similarities in consumer trends as discussed in Section \ref{sec:network}. This enabled us to build a set of $10,238$ businesses. The top $16$ retail categories examined were as follows: Bakeries, Bars, Burger Joints, Clothing Stores, Cocktail Bars, Coffee Shops, Fast Food Restaurants, French Restaurants, Grocery Stores, Hotels, Italian Restaurants, Nightclubs, Pizza Places, Pubs, Sandwich Places, and Supermarkets.

Figure \ref{fig:impact_individual} shows the impact measured for each Burger Joint where a Coffee Shop opened within its spatial neighborhood. We observe a heterogeneous distribution of impact scores. However $66.1\%$ of the $56$ burger joints in this configuration were impacted negatively. The median impact is $0.85$ which corresponds to a \emph{$15\%$ decrease in demand to Burger Joint when a Coffee Shop opens within 100 meters}. In this analysis, due to the skewness of the impact measures distribution, we use the median and the percentage of positively (or negatively) impacted businesses as key numbers to represent the directed impact of pairs of categories. Furthermore, we note the structure of Burger Joints in Figure \ref{fig:impact_individual}. It is composed of two parts: a high impact peak followed by a linear decline. Interestingly, this structure is consistent across all pairs of categories, generalized with a peak of $\sim 10\%$ of high impacts shops ($I > 1.5$), followed by a linear decline of impact ($\sim 80-90\%$ of shops), and succeeded with an optional $\sim 10\%$ discontinuously negatively impacted shops, representing competition in the area ($0 < I < 0.5$).

In Figure \ref{fig:impact_matrix} we show the matrix of median impact of our top $16$ retail categories on each other. A few observations are worth noting. First, impact scores on the diagonal tend to be negative. This observation shows and quantifies the direct competitive effect of businesses belonging to the same category. Second, most impact scores are negative suggesting the general tendency for retail establishments to compete. 
Third, some cooperative pairs are noticeable (e.g. Hotels on Pizza Place: $70\%$ of positive impact, median impact: 1.2; Nightclubs on French Restaurants: $67\%$ of positive impact, median impact: $1.8$). We also note that certain network links seen in Figure~\ref{fig:networkViz} between categories can be found as positive impact pairs in the matrix, such as Nightclubs and French Restaurants and Fast Food Restaurants and Bars.


\subsubsection{Building a competitive ranking} 
Using the aforementioned results we list a ranking of homogeneous competition in Table \ref{tab:competitiveRanking}. This ranking shows the top categories in terms of negative impact, when a business of the same category opens within $100$ meters. The list suggests that fast food restaurants feature the strongest homogeneous negative impact ($-21\%$), closely followed by other businesses where people may visit for food or a snack, namely Bakeries, Pizza Places, Coffee Shops, Sandwich places. Note also the high probability of negative impact for Bakeries with a $73.1\%$ chance to be negatively impacted.

Our methodology can also be applied to heterogeneous pairs of businesses to determine which new business type would create the greatest competition or cooperation for an existing business. Using Coffee Shops as an example, we saw in Table~\ref{tab:competitiveRanking} that a new Coffee Shop creates a median impact of $I=0.84$. We similarly see in Figure \ref{fig:impact_matrix} a competitive effect from new Burger Joints, which result in a median impact of $I=0.85$. A competitive ranking of this type can be established for all categories to better understand trends in the impact of new businesses.

To our knowledge the methodology described above is novel, and despite the limitations that we discuss in more detail in Section~\ref{sec:conclusion}, it allows for the principled quantification of the impact that new businesses have in the retail ecosystem they operate. We exploit this formulation in the context of a prediction task in Section~\ref{sec:ml} where impact scores are modeled as target labels and the goal becomes to predict the impact of new businesses opening on their local environment. From an economic perspective these results highlight the most competitive types of business categories that operate in cities and moreover the highlight cases of categories where cooperation in terms of customer flows is more likely to emerge.

\subsubsection{Takeaways}
Our approach can enable a quantitative metric to measure competitive and cooperative behaviors between pairs of categories. Further, it highlights general trends, such as competition between food-related categories and cooperation between certain clusters of categories. However, reducing the complexity has two limitations. It limits the number of data as it filters out neighborhoods with more than one business opening within $W_T$. It also over-simplifies the problem. 
Often, for a given pair of categories, there is heterogeneity in the impact measure between those two categories (e.g. there is a range in the impact of a new Coffee Shop on a Bakery). This suggests there are complex interactions between a new venue and its environment and there may be additional factors to consider. In the following sections, we study the impact of combinations of multiple new businesses coupled with the network topology of their environment. This methodology leads to insights which can help determine the optimal location for a new venue. 


\section{Predicting New Business Impact}
\label{sec:ml}
\begin{table*}
\centering
\captionsetup{justification=centering}
\makebox[\textwidth][c]{\begin{tabular}{M{3.6cm}|c|c|c|c|c|c}
\cline{1-7}
Category & Coffee Shops & Bars & Italian Restaurants & Hotels & French Restaurants & Bakeries \\ \cline{1-7}
Train/Test Size & 1956/218 & 2275/253 & 1629/181 & 2129/237 & 1631/181 & 1526/170  \\ 
\cline{1-7}
AUC (Business Features Baseline) & 0.611 & 0.603 & 0.618 & 0.621 & 0.701 & 0.771  \\ 
\cline{1-7}
AUC (Network Features Baseline) & 0.549 & 0.551 & 0.564 & 0.557 & 0.591 & 0.584  \\ 
\cline{1-7}
\textbf{AUC (All Features)} & \textbf{0.627} & \textbf{0.642} & \textbf{0.658} & \textbf{0.702} & \textbf{0.742} & \textbf{0.861}  \\
\cline{1-7}
\end{tabular}}
\caption{AUC Scores for a subset of categories using a Gradient Boosting model.}
\label{tab:mlresults}
\end{table*}


In Section~\ref{sec:firstdegree}, we explored the impact of the opening of a single business on its neighborhood. Next, we investigate in the form of a prediction task the cumulative impact of multiple shops opening within the same spatial and temporal neighborhoods, as defined in the previous section. 

\subsection{Prediction Task} 
To illustrate our methodology, let us consider the example of French Restaurants. To understand the impact of new venues on French Restaurants in Europe and North America, we consider all new venues that opened within the spatial radius $r_s$ of a French restaurant within a given period $r_t$. We then examine how the demand of that given French restaurant changed after the opening of the new venues. We aim to predict whether the opening of those new venues will have a positive or negative impact on the French restaurant given network features about the venue, and a count of the types of venues that opened within $r_t$. We consider this prediction task for all retail categories aiming to understand how each category is impacted by new venues opening nearby. 

We model the impact of new venues as a binary classification task where the impact on the existing venue is the dependent variable and the features described below in Section~\ref{sec:features} are independent variables. We further represent a positive impact label as 1 and a negative impact label with a 0. Considering a supervised learning methodogy, our goal then becomes to learn an association of the input feature vector $\mathbf{x}$ with a binary label $y$. We experiment with a number of supervised learning algorithms described in the following paragraphs. All network features were min-max normalised. We split our dataset into training and test sets with the training set consisting of 80\% of the data and the test set consisting of the remaining 20\%. We perform 5-fold cross-validation to pick the best performing model and report the subsequent accuracy of prediction. Finally, we also sub-sample our dataset performing our predictions on a balanced dataset of positive and negative classes. 
 
\subsection{Extracting features}
\label{sec:features}
\subsubsection{Business features}\label{business_section}
When a new venue $v_n$ opens at time $t_{v_n}$ within a given spatial neighborhood $SN(v_n,r_s)$ we count all other businesses which opened within the temporal neighborhood $TN(v_n,r_t)$ of $v_n$ and in the same spatial neighborhood $SN(v_n,r_s)$. For each existing venue $v_e$, the counts of each category of venue within its spatial and temporal neighborhood is encoded as a feature. We refer to these features as \emph{Business features}.

\subsubsection{Network features}\label{network_section}
As described above, we utilize network topology measures at the venue neighborhood level considering each venue $v_e \in SN(v_n,r_s)$ where each venue is a node in the network. Edges in the network represent the number of transitions from the source venue to the destination venue. These features are described as follows. The \emph{in-degree} of the venue $v_e$ is the number of edges coming from the other venues to $v_e$. The \emph{out-degree} of the venue $v_e$ is the number of edges coming from $v_e$ to the other venues. The \emph{degree} of $v_e$ is the total number of edges between $v_e$ and the other venues. The \emph{closeness centrality} of a venue $v_e$ is the average of its shortest paths to all the other venues of the network. The \emph{diversity} of a venue $v_e$ is defined by the Shannon equitability index from information theory which calculates the variety of the neighbors of venue $v_e$ \cite{Shannon}. 
These features give a representation of how well a given venue is connected to the rest of the city network and also describe the distribution of its customers. 

\subsection{Evaluation}  
In this section, we report our findings on the predictive ability of the features as well as the supervised learning models we employ in the prediction task. We compare the predictions against different baselines:
\begin{itemize}
    \item Network connectivity features, as described in section \ref{business_section}.
    \item Business features, as described in section \ref{network_section}.
\end{itemize}
We first discuss how our combined model outperforms both baselines, then show the predictive capabilities of different categories, and lastly discuss the value of network metrics in our prediction task.
 
\textbf{Model Selection:} 
We explored a number of different models, including Logistic Regression, Gradient Boosting, Support Vector Machines, Random Forests, and Neural Networks. As described above, we train our models to predict the impact of new businesses on a given type of retail category.
When aggregating our data across all categories and working to predict the impact of a new venue on any type of existing venue category, this resulted in models with low predictive power. As discussed previously in Section~\ref{sec:firstdegree}, this suggests there is significant heterogeneity of behavior across different category types and therefore a training strategy tailored for each category because more effective. To take this into account, we segregated our data by category and built a separate model for each type of category. We individually modelled the ten retail categories with the largest total number of data points. These categories were as follows: Bars, Hotels, Coffee Shops, Italian Restaurants, French Restaurants, Bakeries, Clothing Stores, Pizza Places, Nightclubs, and Grocery Stores. To compare our supervised learning models, we calculated the average AUC scores across all ten categories using our different feature baselines.
 
Our combined models, using both business and network features, had a resulting AUC of 0.611 for Logistic Regression, 0.687 for Gradient Boosting,  0.631 for Support Vector Machines, 0.640 for Random Forest, and 0.667 for Neural Networks. These results suggest that venue-specific metrics can support the prediction of the impact of a new venue. Across all categories, we saw that Gradient Boosting was the most robust model for our task with the highest average AUC across all ten categories. As such, we use Gradient Boosting to further analyze our models below. 

\textbf{Variations across Categories}
We next examine the predictability of different category types. Table~\ref{tab:mlresults} shows the predictability of six of the ten categories trained using a Gradient Boosting model. The AUCs of the remaining categories are as follows: Clothing Stores (AUC: 0.664, train/test size: 1362/152), Pizza Places (AUC: 0.682, train/test size: 1287/143), Nightclubs (AUC: 0.686, train/test size: 622/70), and Grocery Stores (AUC: 0.690, train/test size: 581/65). We see a range in the predictability of different venue types: Bakeries have the highest AUC of 0.861 and Coffee Shops have the lowest AUC of 0.627. This suggests that certain activities adapt to a more heterogeneous set of environment, making them more challenging to predict. These more challenging categories tend to be those that are ubiquitous and general, such as Bars and Coffee Shops.
 
\begin{figure} 
\begin{center}
\includegraphics[width=8.5cm]{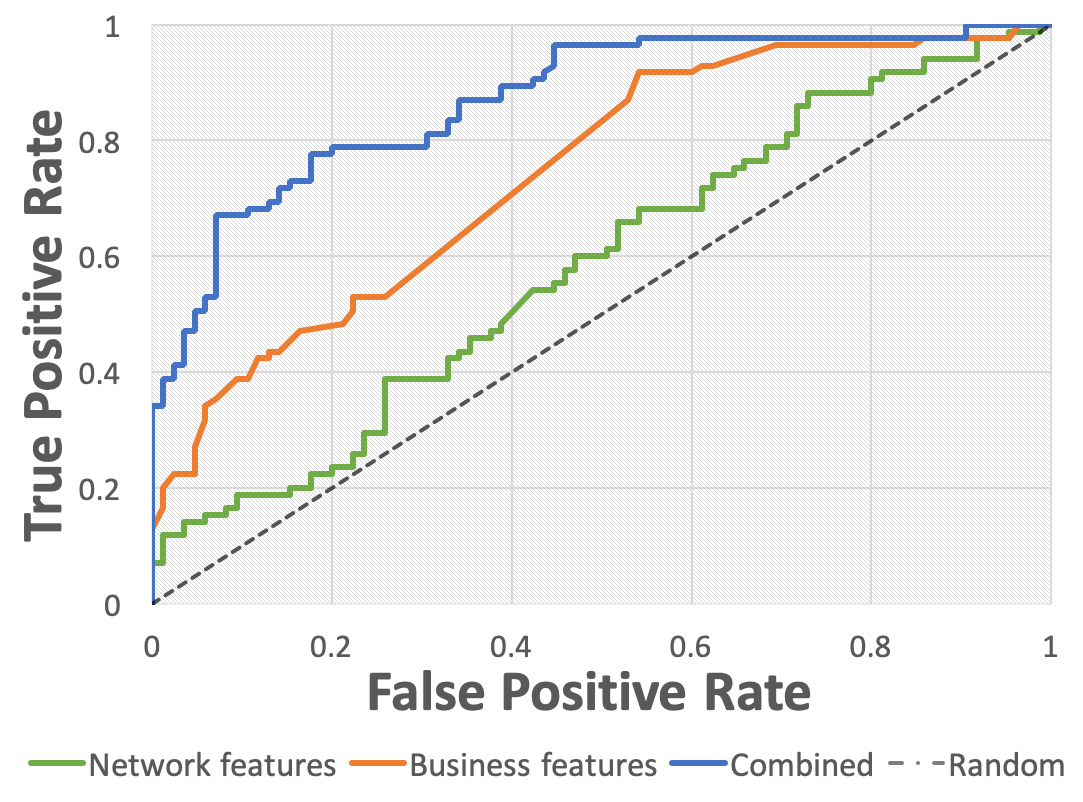}
\caption{ROC Curves of Bakeries for the performance of each class of features and for the combined model.}
\label{fig:varyingFeatures}
\end{center}
\end{figure}

\textbf{The Value of Network Metrics:} To understand the influence of different feature classes on our results, we run our Gradient Boosting model on each class separately and in combination. We report our observations in Figure~\ref{fig:varyingFeatures} and note that for Bakeries the category features alone reach AUC of $\approx$ 0.77, the network features alone reach AUC of $\approx$  0.60 and combined model leads to a higher overall AUC of $\approx$ 0.85. These results, which were consistent across all retail categories, suggest using network features in lieu of venue features alone supports our prediction task. We see that the two feature classes together have the best prediction results, suggesting that the impact of new venues is driven by a number of forces, including the network connectivity, which modulate the nature of their interactions within their neighborhood.
 
\section{Conclusion And Future Work}
\label{sec:conclusion}
Our methodology highlights the power of complex networks measures in building machine learning prediction models. This is especially valuable for systems in which interactions between agents must be taken into account.
We began in Section~\ref{sec:network} of this work by detecting patterns of cooperation in urban networks and quantifying similarities and differences in network structure across cities. Next, in Section~\ref{sec:firstdegree} we measured the impact of new businesses isolating cases in which exactly one new venue opened within a given region. Using urban network topology measure and our business impact metrics, we developed a machine learning model to predict the optimal location for a new business in Section~\ref{sec:ml}. The novelty of our approach in methodological terms stems from the use of complex networks measures, combined with machine learning methods to tackle modern societal challenges.
These results can support policy makers, business owners, and urban planners as they have the potential to pave the way for the development of sophisticated models describing urban neighborhoods and help determine the optimal conditions for establishing a new venue. 

One of the limitations of our work is that we only examine the impact of retail venues. However, this analysis can be more broadly applied to venues of other categories as well. The present study sets the frame for further general studies. As one example of future work, one could examine the impact of new bus stops or transport hubs on those venues within their proximity. Additionally, in this work we conduct a preliminary examination of the variations in network trends across cities worldwide. Future work could expand upon this to explore the duality between general network trends and cultural consumer idiosyncrasies across cities. Our analysis could also be further expanded by considering temporal network analysis, examining the variations in features across different time intervals of interest. 

\section{Acknowledgements}
We thank Foursquare for supporting this research by providing the dataset used in the analysis. \\
Jordan Cambe would like to thank the GdR-ISIS and the ED PHAST (52) for supporting his work in the University of Cambridge. \\
Krittika D'Silva's work was supported through the Gates Cambridge Trust.

\bibliographystyle{abbrv}
\bibliography{main}  

\end{document}